\documentclass[final,3p,times]{elsarticle}

\usepackage{amssymb, amsmath,framed,amsthm}

\journal{Journal of Computational Physics {\rm or} Computer Physics Communications}

\usepackage{color,soul}
\usepackage{graphicx}

\usepackage[normalem]{ulem}

\usepackage[colorlinks=true,linkcolor=blue]{hyperref}
\newcommand{\p}{\partial}
\newcommand{\f}[2]{\frac{#1}{#2}}

\newcommand{\pd}[2]{\frac{\p #1}{\p #2}}

\newcommand{\parl}{\parallel}
\newcommand{\unit}[1]{\hat{\bm{#1}}}

\makeatletter
\newcommand{\vast}{\bBigg@{4}}
\newcommand{\Vast}{\bBigg@{5}}

\newcommand\gs{{\tt GS2}~}
\newcommand{\bm}[1]{{\boldsymbol {\mathrm #1}}} 
\newcommand{\bhat}{\unit{b}}

\newcommand{\vth}{v_{\textnormal{th}}}
\newcommand{\vthe}{v_{\textnormal{th},e}}
\newcommand{\vthi}{v_{\textnormal{th},i}}
\newcommand{\vthr}{v_{\textnormal{th},r}}
\newcommand{\vths}{v_{\textnormal{th},s}}
\newcommand{\rhoi}{\rho_i}
\newcommand{\tld}{\tilde}
\newcommand{\stella}{\texttt{stella}~}
\newcommand{\sstella}{\texttt{stella}}
\newcommand{\beq}{\begin{equation}}
\newcommand{\eeq}{\end{equation}}
\newcommand{\vpa}{v_{\parl}}
\newcommand{\vpas}{v_{\parl}}
\newcommand{\vpe}{v_{\perp}}
\newcommand{\grad}{\nabla}
\newcommand{\mbf}{\mathbf}
\newcommand{\gyroR}[1]{\left<#1\right>_{\mbf{R}}}

\newcommand{\dz}{\Delta z}
\newcommand{\dt}{\Delta t}
\newcommand{\gperp}{\grad_{\perp}}
\newcommand{\phitb}{\varphi}

\newcommand{\vvol}{d^3v}

\newcommand{\nvpa}{N_{v_{\parallel}}}
\newcommand{\defeq}{\doteq}
\newcommand{\gks}{g_{\mbf{k},s}}
\newcommand{\phik}{\phitb_{\mbf{k}}}
\newcommand{\gksnorm}{\tilde{g}_{\mbf{k},s}}
\newcommand{\tnorm}{\tilde{t}}
\newcommand{\vpanorm}{\tilde{v}_{\parl}}
\newcommand{\vpasnorm}{\tilde{v}_{\parl}}

\newcommand{\musnorm}{\tilde{\mu}_s}

\newcommand{\vsnorm}{\tilde{v}_s}
\newcommand{\tempnorm}{\tilde{T}_s}
\newcommand{\nnorm}{\tilde{n}_s}
\newcommand{\gradnorm}{\tilde{\grad}}

\newcommand{\maxsnorm}{\exp\left(-\vsnorm^2\right)}
\newcommand{\bnorm}{\tld{B}}
\newcommand{\phinorm}{\tilde{\phitb}_{\mbf{k}}}
\newcommand{\bess}{J_0(a_{\mbf{k},s})}

\newcommand{\vshift}{v_{\parallel,*}}
\newcommand{\dvpa}{\Delta \vpa}
\newcommand{\tupwnd}{u_t}
\newcommand{\zupwnd}{u_z}
\newcommand{\dblbar}[1]{\overline{\overline{#1}}}
\newcommand{\tldg}{\tld{g}}
\newcommand{\tldphi}{\tld{\phitb}}
\newcommand{\nbar}{\overline{n}}
\newcommand{\nbarbar}{\dblbar{n}}
\makeatother

\DeclareMathAlphabet{\mathpzc}{OT1}{pzc}{m}{it}

\begin{document}

\begin{frontmatter}


\title{\sstella: a mixed implicit-explicit, $\delta f$-gyrokinetic code for general magnetic field configurations}
\author[Oxford,CCFE]{M.~Barnes\corref{cor}}
\ead{michael.barnes@physics.ox.ac.uk}
\cortext[cor]{Corresponding author}

\author[Oxford,CCFE]{F.~I.~Parra}
\author[Maryland]{M.~Landreman}

\address[Oxford]{Rudolf Peierls Centre for Theoretical Physics, University of Oxford, Oxford OX1 3NP, United Kingdom}
\address[CCFE]{Culham Centre for Fusion Energy, Culham Science Centre, Abingdon OX14 3DB, United Kingdom}
\address[Maryland]{Institute for Research in Electronics and Applied Physics, University of Maryland, College Park, Maryland 20742, USA}


\begin{abstract}
Here we present details of a mixed implicit-explicit numerical scheme for the solution of the gyrokinetic-Poisson system of equations in the local limit.  This scheme has been implemented in a new code called \sstella, which is capable of evolving electrostatic fluctuations with full kinetic electron effects and an arbitrary number of ion species in general magnetic geometry.  We demonstrate the advantages of this mixed approach over a fully explicit treatment and provide linear and nonlinear benchmark comparisons for both axisymmetric and non-axisymmetric magnetic equilibria.
\end{abstract}


\end{frontmatter}




\section{Introduction}\label{s:intro}

The turbulent transport of particles, momentum and energy places a fundamental constraint on the confinement -- and thus performance -- of magnetic confinement fusion (MCF) plasmas.  
This turbulence is challenging to simulate for a number of reasons: the collisional mean free path in MCF plasmas is often larger than the system size, necessitating a kinetic treatment; the presence of a strong mean magnetic field makes the turbulence highly anisotropic;  and the characteristic space-time scales of the turbulence are much smaller than the space-time scales associated with the mean density, flow and temperature.  On the face of it, one must thus resolve six-dimensional phase space dynamics involving multiple space-time scales spanning several orders of magnitude.

However, by exploiting the anisotropy of the turbulence and scale separation in space and time, it is possible to reduce the complexity of the problem considerably.  This is the approach taken by $\delta f$-gyrokinetics~\cite{cattoPP78,friemanPoF82}, which we describe in detail in Sec.~\ref{s:model}.  In brief, it eliminates the fast gyro-motion time scale and the gyro-angle phase space variable, and it separates the space-time scales of mean and fluctuating quantities.  The development of gyrokinetics (and the gyrofluid models derived from it) and its subsequent numerical implementation in a wide range of codes~\cite{parkerPRL93,kotschCPC95,linS98,jenkoCPC00,jostPoP01,candyJCP03,idomuraNF03,watanabeNF06,jollietCPC07,peetersCPC09} has facilitated a leap forward in our ability to accurately model, predict and understand turbulent transport in MCF plasmas.  Given the proliferation of gyrokinetic codes and their growing success in describing experimental behavior, it is worth considering if there is a need for yet another gyrokinetic code.

Most of the existing $\delta f$-gyrokinetic codes have been developed to simulate tokamak plasmas, with only a handful~\cite{baumgaertelPoP11,xanthopoulosPRX16,sanchezPPCF13,nunamiPFR10,spongNF17} capable of simulating non-axisymmetric magnetic field configurations. Indeed, there is a relative paucity of simulations for stellarators, and most of them are linear simulations in the local (or flux-tube) limit with Boltzmann electrons.  As it is beneficial to have a diversity of numerical approaches to the same class of problems, we have developed a new code \stella with the goal of enabling routine and efficient simulation of turbulence in stellarators.  The numerical scheme employed in \stella is distinguished from other gyrokinetic codes through its use of a mixed implicit-explicit time advance scheme with strong-stability-preserving methods to maximize the allowable time step size for both linear and nonlinear calculations.  As such, \stella is expected to be a useful tool for simulations of turbulence in both tokamaks and stellarators.

The flux tube approach is very efficient for simulating plasma turbulence in an axisymmetric magnetic field because a single flux tube spanning a $2\pi$ domain in poloidal angle effectively samples an entire flux surface.  For a non-axisymmetric confining field, this is not the case: Because a flux tube is asymptotically small in the local limit, it in principle must extend infinitely far along a magnetic field line as it ergodically samples an entire flux surface.  As this is not feasible, one must either be content with sampling a fraction of a flux surface or one must go beyond the flux tube approach to simulate a flux annulus encompassing a full flux surface.  To our knowledge only a few codes~\cite{xanthopoulosPRL14,sanchezPPCF13,spongNF17} currently allow for the simulation of an entire flux surface, and results of this type are few and far between.  While the version of \stella documented here employs the flux tube approximation, the numerical scheme has been formulated with the aim of extending the code to treat the full flux annulus.  This is discussed in more detail when the algorithm is introduced in Sec.~\ref{s:algorithm}.

The paper is organized as follows:  In Sec.~\ref{s:model} we describe the gyrokinetic model and state the governing equations.  We then give an overview of the coordinates used in \stella and the options available for specifying magnetic geometry in Sec.~\ref{s:coordinates}.  The normalized simulation equations are provided in Sec.~\ref{s:normalization} before detailing the numerical scheme employed by \stella in Sec.~\ref{s:algorithm}.  We then compare simulation results from \stella with those from the widely-benchmarked gyrokinetic code \gs in Sec.~\ref{s:benchmarks} before summarizing and discussing possibilities for future work on \stella in Sec.~\ref{s:futurework}.

\section{Model equations}\label{s:model}

Derivations of the $\delta f$ gyrokinetic model employed by \stella are abundant in the plasma physics literature (cf.~\cite{cattoPP78,friemanPoF82,brizardRMP07,parraPPCF08,abelRPP13}) and so we provide only a brief overview of its orderings and assumptions here.
The essence of gyrokinetics is an assumption that all dynamics of interest are slow compared to particle gyration about a mean magnetic field.  This allows one to usefully split particle motion into a rapid, approximately circular orbit about the magnetic field and the movement of this orbit's centre, called the guiding centre; i.e., particle position $\mbf{r}$ is given by $\mbf{r}=\mbf{R}+\bm{\rho}$, with $\mbf{R}$ the guiding centre position, and $\bm{\rho}=\unit{b}\times\mbf{v}/\Omega$ the gyroradius vector.  Here, $\unit{b}$ is the unit vector pointing along the mean magnetic field, $\mbf{v}$ is the particle velocity, and $\Omega=ZeB/mc$ is the frequency of gyration about the mean field, with $Z$ charge number, $e$ proton charge, $B$ magnetic field strength, $m$ particle mass, and $c$ speed of light.  After making this split, one averages over the rapid gyration to eliminate the gyration angle as a phase space variable.  Gyrokinetics thus describes the motion of charged rings as they stream along the mean magnetic field and slowly drift across it.

We restrict our attention to plasmas with sub-sonic mean flows and electrostatic fluctuations so that to lowest order the electric field $\mbf{E}=-\grad\phitb$, with $\phitb$ the fluctuating electrostatic potential.  These constraints are not required by the gyrokinetic model and can thus be relaxed in future work.  The use of $\delta f$ gyrokinetics does rely on a separation of space-time scales between
the plasma equilibrium and the turbulent fluctuations.  In particular, the particle distribution function for species \textit{s}, denoted $f_s$, is expressed as $f_s=F_s+\delta f_s$, and the following orderings are imposed:
\beq
\frac{\delta f_s}{f_s} \sim \frac{\omega_s}{\Omega_s} \sim \frac{\rho_s}{L} \sim  \frac{k_{\parallel}}{k_{\perp}} \sim k_{\parallel}\rho_s \sim \frac{e\phitb}{T_s}\sim  \epsilon \ll 1,
\eeq
where $\epsilon$ is the fundamental ordering parameter in gyrokinetics, $\omega$ is a characteristic fluctuation frequency, $\rho=\vth/\Omega$ is thermal gyroradius, $\vth=\sqrt{2T/m}$ is thermal speed, $T$ is temperature, $L$ is a characteristic length associated with the equilibrium, and $k_{\parallel}$ and $k_{\perp}$ are characteristic fluctuation wave numbers along and across the mean magnetic field.

Upon gyro-averaging the Vlasov equation and applying the gyrokinetic ordering given above, one obtains the lowest-order, electrostatic gyrokinetic equation for the distribution of guiding centres $g_s(\mbf{R},\vpa,\mu,t)\defeq \gyroR{\delta f}$:
\beq
\begin{split}
\pd{g_s}{t}&+\vpa\bhat\cdot\grad z\left(\pd{g_s}{z}+\f{Z_se}{T_s}\pd{\gyroR{\phitb}}{z}F_{s}\right) 
- \f{\mu_s}{m_s} \unit{b}\cdot\grad B \pd{g_s}{\vpa} 
+\mbf{v}_{Ms}\cdot\left(\gperp g_s + \f{Z_s e}{T_s} \gperp \gyroR{\phitb}F_{s}\right) \\
&+ \gyroR{\mbf{v}_{E}}\cdot \nabla_{\perp} g_s
+\gyroR{\mbf{v}_{E}}\cdot\grad\big|_E F_{s}= 0,
\end{split}
\label{eqn:gke}
\eeq
where, unless noted otherwise, derivatives are taken at fixed guiding centre position $\mbf{R}$, parallel speed $\vpa$, and magnetic moment $\mu=m_s v_{\perp}^2/2B$.  Here, $E=m_sv^2/2$ is kinetic energy, $t$ is time, $z$ is a coordinate that measures distance along the magnetic field, $F$ is the mean distribution function (taken here to be a Maxwellian in $E$), $\gyroR{.}$ denotes a gyro-average at fixed $\mbf{R}$,
$\mbf{v}_{Ms}$ is the drift velocity due to the magnetic field gradient and curvature, given by
\beq
\mbf{v}_{Ms} = \frac{\unit{b}}{\Omega_s} \times \left(\frac{\vpe^2}{2}\frac{\nabla B}{B} + \vpa^2\bm{\kappa} \right),
\eeq
$\bm{\kappa}=\unit{b}\cdot\grad\unit{b}$ is the curvature vector, and $\mbf{v}_{E}$ contains the lowest order, fluctuating $\mbf{E}\times\mbf{B}$ drift velocity, given by
\beq
\mbf{v}_{E} = \frac{c}{B}\unit{b}\times \gperp \phitb.
\eeq
The system is closed by coupling to Poisson's equation.  In the usual gyrokinetic ordering -- in which the Debye length is taken to be much smaller than the electron gyroradius -- this reduces to quasineutrality:
\beq
\sum_s Z_s \delta n_s = \sum_s Z_s \int \vvol \left(g_s + \frac{Z_se}{T_s}F_{s}\left(\gyroR{\phitb}-\phitb\right)\right) = 0.
\label{eqn:QN}
\eeq


\section{Coordinates and magnetic geometry}\label{s:coordinates}

The coordinates used in \stella are $(x,y,z,\vpa,\mu)$, with $\vpa=\unit{b}\cdot \mbf{v}$ the speed along the magnetic field, $\mu=m_sv_{\perp}^2/2B$ the magnetic moment, $z$ a coordinate measuring distance along the magnetic field, and $(x,y)$ are coordinates in the plane perpendicular to $\unit{b}$.  The magnetic field vector $\mbf{B}$ is expressed
\beq
\mbf{B}=\grad\alpha\times\grad\psi,
\eeq
where $\alpha$ labels field lines and $\psi$ labels flux surfaces.  The coordinates $(x,y)$ are related
to $\psi$ and $\alpha$ via
\beq
x= \frac{dx}{d\psi}\left(\psi-\psi_0\right)
\eeq
and
\beq
y=\frac{dy}{d\alpha}\left(\alpha-\alpha_0\right),
\eeq
with $\psi_0$ and $\alpha_0$ the values of $\psi$ and $\alpha$ at the centre of the simulation domain.
Note that there is flexibility in defining $x$, $y$, and $z$ within \sstella: details on the currently supported options are given in the subsection on magnetic geometry below.

\subsection{Velocity space grids and integrals}\label{s:vgrids}

In terms of our chosen $(\vpa,\mu,\sigma)$ coordinates, with $\sigma$ the gyration angle, velocity space integrals are of the form
\beq
\int \vvol \ f= \int_0^{2\pi} d\sigma \int_{-\infty}^{\infty}d\vpa\int_{0}^{\infty} d\mu \frac{B}{m} f
\eeq
We truncate the $\vpa$ integral at the cutoff values $\pm v_{\parallel,c}$, with $v_{\parallel,c}$ chosen to ensure that the integrand is sufficiently small for $\vpa > v_{\parallel,c}$.  The $\vpa$ grid points are then chosen to be equally spaced on the interval $[-v_{\parallel,c},v_{\parallel,c}]$.  The number of parallel velocities $\nvpa$ is constrained in \stella to be even so that $\vpa=0$ is not included in the grid.  This choice facilitates parallelization and avoids the need to apply a special treatment to phase space points where $\vpa=\bhat\cdot\grad B=0$, which decouple from all other points for a collisionless plasma.
The $\vpa$ integral is approximated numerically using an average of Simpson's $3/8$ rule and composite Simpson's rule at the final four points at either end of the $\vpa$ domain and pure composite Simpson's rule at the interior points (cf.~\cite{burden}).

The $\mu$ grid points are chosen according to Gauss-Laguerre quadrature (cf.~\cite{abramowitzStegun}):
\beq
\begin{split}
\int_{0}^{\infty} d\mu \frac{B}{m} f &= \int_{0}^{\infty} d\left(\frac{\mu B_0}{T}\right) \frac{TB}{mB_0} \exp(-\mu B_0/T) f \exp(\mu B_0/T) \\
&\approx \sum_{i=1}^{N_{\mu}} w_i f(\hat{\mu}_i) \exp(\hat{\mu}_i) \frac{TB}{mB_0},
\end{split}
\eeq
with $N_{\mu}$ the number of $\mu$ grid points, $\hat{\mu}=\mu B_0/T$, and $B_0$ is a free parameter.  It is desirable to choose $B_0$ to be independent of $z$, as otherwise the physical $\mu$ grid would be $z$-dependent and would complicate computation of $z$ derivatives at fixed $\mu$; i.e.,
\beq
\left(\pd{}{z}\right)_{\mu} = \left(\pd{}{z}\right)_{\hat{\mu}}  + \left(\pd{\hat{\mu}}{z}\right)_{\mu} \left(\pd{}{\hat{\mu}}\right)_z = \left(\pd{}{z}\right)_{\hat{\mu}} + \frac{\mu}{T}\pd{B_0}{z}\left(\pd{}{\hat{\mu}}\right)_z.
\eeq

To satisfy the boundary condition $f(\vpe\rightarrow\infty)\rightarrow0$, we must choose our maximum $\vpe$ at each $z$ so that $f$ evaluated there is approximately zero.  Denoting $v_{\perp,c}$ as the smallest acceptable value satisfying $f(v_{\perp,c})\approx 0$, we obtain the following constraint:
\beq
\hat{\mu}_{\textnormal{max}} \geq \frac{v_{\perp,c}^2}{\vths^2}\f{B_0}{B_{\textnormal{min}}},
\eeq
where $B_{\textnormal{min}}$ is fixed by the magnetic geometry, $\hat{\mu}_{\textnormal{max}}$ is fixed by the choice of $N_{\mu}$, and $v_{\perp,c}/\vths$ is an input parameter.  To ameliorate the CFL constraint on the time step size, we want to minimize the largest value of $\vpe$, and thus $\hat{\mu}_{\textnormal{max}}$, included in the simulation.  Combined with the above inequality, we find
\beq
B_0 = B_{\textnormal{min}} \frac{\vths^2}{v_{\perp,c}^2}\hat{\mu}_{\textnormal{max}}.
\eeq

\subsection{Real space grids and boundary conditions}\label{s:realgrids}

Periodic boundary conditions are enforced in $x$ and $y$ by expressing the guiding centre distribution $g$ in terms of Fourier harmonics:
\beq
g(x,y,z,\vpa,\mu,t) \defeq \sum_{k_x,k_y}\hat{g}_{\mbf{k}}(z,\vpa,\mu,t)\exp(ik_x x + ik_y y),
\eeq
where $\hat{g}_{\mbf{k}}=\hat{g}_{k_x,k_y}$.
This is justified as long as $k_{x} \sim k_y \gg 1/L$, so that turbulence at separate ends of the $(x,y)$ domain is decorrelated and thus statistically periodic.  This local, or flux tube, approximation
is routinely used to model micro-instabilities and turbulence in axisymmetric magnetic field configurations and has been successfully validated across a range of experiments (cf.~\cite{whitePoP13,citrinNF14,gorlerPoP14,howardPoP16,nakataNF16,vanwykPPCF17}).  For
non-axisymmetric field configurations, a single flux tube would in principle need to extend infinitely far along the magnetic field line as it ergodically samples a flux surface.  A version of \stella that simulates a flux annulus encompassing a full flux surface is currently in development to address this deficiency.

The grid in the parallel coordinate $z$ is equally spaced, with the number of $z$ points $N_z$ forced to be odd in order to guarantee points at $z=\pm \pi$ and $z=0$.  The boundary condition in $z$ is a generalization of the `twist and shift' boundary condition~\cite{beerPoP95} in which different $k_x$ values are coupled at the boundaries of the $z$-domain.  For the sake of definiteness, we choose here $z=\zeta$ and $\alpha = \theta-\iota\zeta$, with $\theta$ and $\zeta$ straight-field-line poloidal and toroidal angles, respectively, and $\iota=\iota(\psi)$ the rotational transform.  For an arbitrary fluctuating quantity $A$, physical periodicity dictates $A(x,y(\theta,z),z)=A(x,y(\theta,z+2\pi p),z+2\pi p)$, where $p=M/N_p$ with
$M$ any integer and $N_p$ defined so that the magnetic geometry is 
periodic in $\zeta$ with period $2\pi/N_p$.
In terms of the spectral representation, this parallel boundary condition becomes
\beq
\sum_{k_x,k_y} \hat{A}_{k_x,k_y}(z)\exp\left(ik_{y}y(\theta,z)+ik_{x}x\right) = \sum_{k_x,k_y} \hat{A}_{k_x,k_y}(z+2\pi p)\exp\left(ik_{y}y(\theta,z)+ i\left(k_x - \delta k\right)x\right)\exp\left(-2\pi p i k_y \frac{dy}{d\alpha}\iota(\psi_0) \right),
\eeq
with $\psi_0$ the value of $\psi$ on the flux surface of interest and $\delta k =2\pi p k_y (dy/d\alpha)(d\psi/dx) (d\iota/d\psi)_{\psi_0}$.  Orthogonality of the Fourier harmonics then implies that
\beq
\hat{A}_{k_x,k_y}(z) = \hat{A}_{k_x',k_y}(z+2\pi p)\exp\left(-2\pi p i k_y \frac{dy}{d\alpha}\iota(\psi_0)\right),
\eeq
with $k_x'=k_x + \delta k$.  An outgoing boundary condition is applied at the end of each set of connected $2\pi$ segments, with a zero incoming boundary condition on $g$.

\subsection{Magnetic geometry}\label{s:geometry}

With these coordinate choices, there are eight independent geometrical quantities appearing in the gyrokinetic equation: $B$, $\unit{b}\cdot\grad z$, $|\grad x|^2$, $|\grad y|^2$, $|\grad x \cdot \grad y|$, $(\unit{b}\times \grad B) \cdot \grad y$, $(\unit{b}\times \bm{\kappa}) \cdot\grad x$, and $(\unit{b}\times\bm{\kappa})\cdot\grad y$.  There are currently two options in \stella for obtaining these quantities.  The first option is to use a magnetic equilibrium generated by $\texttt{VMEC}$~\cite{hirshmanPoF83,hirshmanCPC86}.  A module in \stella takes the $\texttt{VMEC}$ output and computes all of the geometrical quantities needed to solve the gyrokinetic equation on a user-selected flux surface and field line.  For this option, the $x$ and $y$ coordinates are chosen so that $dx/d\psi=\sqrt{\psi_{LCFS}/\psi}/aB_r$ and $dy/d\alpha=a\sqrt{\psi/\psi_{LCFS}}$, with $a$ the effective minor radius computed by $\texttt{VMEC}$, $B_r=2\psi_{LCFS}/a^2$, $\psi=\psi_t$ the enclosed toroidal flux divided by $2\pi$, and $\psi_{LCFS}$ its value at the outermost flux surface computed by $\texttt{VMEC}$ (conventionally called the `last closed flux surface').

The second option, valid only for axisymmetric magnetic field configurations, is to specify a set of Miller parameters that are used to construct a local solution to the Grad-Shafranov equation~\cite{millerPoP98}.  In brief, the cylindrical coordinates ($R,Z$) of the desired flux surface are assumed to be of the form
\begin{gather}
R(r,\vartheta) = R_0(r) + r \cos(\vartheta + \sin\vartheta \arcsin \delta(r)), \\
Z(r,\vartheta) = \kappa(r) r\sin(\vartheta),
\end{gather}
where the flux label $r$ is the half-diameter of the flux surface at the height of the magnetic axis, $R_0$ is the average of the minimum and maximum values of the major radius at the height of the magnetic axis, and $\vartheta$ is a poloidal angle.  Note that $\vartheta$ is not in general a straight-field-line angle.  With this assumed form for the desired flux surface (and for neighboring surfaces), all of the required geometric quantities can be calculated by providing: the flux surface location $r$, the local major radius $R_0$ and its derivative $dR_0/dr$, the local elongation $\kappa$ and its derivative $d\kappa/dr$, the local triangularity $\delta$ and its derivative $d\delta/dr$, the local safety factor $q$ and its derivative $dq/dr$, and the MHD $\alpha$-parameter $(4\pi p_{\textnormal{tot}}/B_r^2)( d\ln p_{\textnormal{tot}} / dr)$, where $p_{\textnormal{tot}}$ is the species-summed plasma pressure.  For this option, the $x$ and $y$ coordinates are chosen so that $dx/d\psi = q/rB_r$ and $dy/d\alpha=(d\psi/dr)/B_r$, with reference length $a$ the half-diameter of the plasma volume at the height of the magnetic axis, $B_r$ is the user-specified reference magnetic field strength, and $\psi$ is the enclosed poloidal flux divided by $2\pi$.  


\section{Simulation equations}\label{s:normalization}

The gyrokinetic equation solved in \stella is obtained by taking the discrete Fourier transform $\mathcal{F}_{\mbf{k}}$ of the gyrokinetic equation~(\ref{eqn:gke}) in $x$ and $y$ and multiplying by the normalizing factor
$(a^2/\rho_r \vthr)\exp(-v^2/\vths^2)/F_{s}$, with $\vthr\defeq\sqrt{2T_r/m_r}$, $\rho_r\defeq\vthr/\Omega_r$, and $\Omega_r\defeq eB_r/m_rc$.  The reference mass $m_r$, density $n_r$, and temperature $T_r$ are user-specified, while the reference length $a$ and magnetic field strength $B_r$ (different from $B_0$) are determined by the choice of magnetic geometry model as detailed in Subsection~\ref{s:geometry}.  With this choice of normalization, the gyrokinetic equation solved by \stella is
\beq
\begin{split}
\pd{\gksnorm}{\tnorm}&
+\frac{\vths}{\vthr}\vpasnorm\bhat\cdot\gradnorm z\left(\pd{\gksnorm}{z}+\frac{Z_s}{\tempnorm}\pd{\bess\phinorm}{z} \maxsnorm \right) 
- \frac{\vths}{\vthr}\musnorm \bhat \cdot \gradnorm \bnorm \pd{\gksnorm}{\vpasnorm}
 \\
&+ i \omega_{d,\mbf{k},s}\left(\gksnorm + \frac{Z_s}{\tempnorm} \bess \phinorm \maxsnorm \right) +i \omega_{*,\mbf{k},s} \bess \phinorm  + \mathcal{N}_{\mbf{k},s} = 0,
\end{split}
\label{eqn:gkstella}
\eeq
where 
$\tnorm\doteq t a/\vthr$, $\tempnorm\doteq T_s/T_r$, $\vpasnorm=\vpas/\vths$, $\gradnorm\doteq a\grad$, $J_0$ is a Bessel function of the first kind, $a_{\mbf{k},s}\doteq k_{\perp}v_{\perp}/\Omega_s$, $k_{\perp}^2=k_x^2|\grad x|^2 + k_y^2|\grad y|^2+2k_xk_y\grad x\cdot\grad y$, $\musnorm\doteq \mu_s B_r/2T_s$, $\bnorm\defeq B/B_r$,
\beq
\gksnorm \doteq \frac{\hat{g}_{\mbf{k},s}}{F_{s}}\exp\left(-\vsnorm^2\right)\frac{a}{\rho_r},
\eeq
\beq
\phinorm \doteq \frac{e\hat{\phitb}_{\mbf{k}}}{T_r}\frac{a}{\rho_r},
\eeq
\beq
\omega_{*,\mbf{k},s} \defeq \frac{k_y\rho_r}{2}aB_r \frac{dy}{d\alpha} \maxsnorm \frac{d\ln F_{s}}{d\psi},
\eeq
\beq
\mathcal{N}_k \doteq \frac{B_r}{2}\frac{dy}{d\alpha}\frac{dx}{d\psi}\mathcal{F}_{\mbf{k}}\left[\mathcal{F}_{\mbf{k}}^{-1}\left[ik_y\rho_r \bess \phinorm\right] \mathcal{F}_{\mbf{k}}^{-1}\left[ik_x\rho_r\gksnorm\right]-\mathcal{F}_{\mbf{k}}^{-1}\left[ik_x\rho_r\bess \phinorm\right]\mathcal{F}_{\mbf{k}}^{-1}\left[ik_y\rho_r\gksnorm\right]\right],
\eeq
and
\beq
\omega_{d,\mbf{k},s} \defeq \frac{\tempnorm}{Z_s \bnorm}\left(\vpasnorm^2\mbf{v}_{\kappa}+\musnorm\mbf{v}_{\grad B}\right)\cdot \left(k_y\rho_r\grad y + k_x\rho_r\grad x\right),
\eeq
with $\mbf{v}_{\kappa} = \unit{b}\times \left(\unit{b}\cdot\gradnorm\unit{b} \right)$ and
$\mbf{v}_{\grad B} = \unit{b}\times \gradnorm\bnorm$.

The normalized form of quasineutrality for Fourier component $\mbf{k}$ is
\beq
\sum_s Z_s \frac{\delta n_{\mbf{k},s}}{n_r} \frac{a}{\rho_r}=\sum_s Z_s\nnorm \left( \frac{2\bnorm}{\pi^{1/2}}\int_{-\infty}^{\infty}d\vpasnorm \int_{0}^{\infty} d\musnorm\bess\gksnorm + \frac{Z_s}{\tempnorm}\left(\Gamma_0(b_{\mbf{k},s})-1\right)\phinorm\right) = 0,
\label{eqn:QNstella}
\eeq
with $\nnorm\defeq n_s/n_r$, $b_{\mbf{k},s}\defeq k_{\perp}^2\rho_s^2/2$, $\Gamma_0(b)\defeq \exp(-b)I_0(b)$, and $I_0$ is a modified Bessel function of the first kind.


\section{Algorithm}\label{s:algorithm}

When electron dynamics are retained in the gyrokinetic equation~(\ref{eqn:gkstella}), the parallel streaming and acceleration terms are scaled up compared to all other terms by a factor of $\vthe/\vthi\sim 60$.
This places a significant restriction on the time step size for explicit time advance schemes.
The severity of this restriction becomes prohibitive at long wavelengths~\cite{wwleeJCP87,candyJCP03}, since the electrostatic potential obtained from quasineutrality~(\ref{eqn:QNstella}) is derived from a polarization density that vanishes at infinite wavelength.  In the absence of electromagnetic effects, this leads to a discretization-dependent CFL condition that scales as either $1/k_{\perp}$ or $1/k_{\perp}^2$, as we demonstrate numerically in the next section.

In light of these considerations, it is desirable to treat the parallel streaming and acceleration terms implicitly.  We achieve this without undue computational expense by employing operator splitting to separate the faster time scales associated with streaming and acceleration from the rest of the dynamics.  We start by splitting the gyrokinetic equation into three pieces:
\beq
\pd{\gksnorm}{t} = \left(\pd{\gksnorm}{t}\right)_{1} + \left(\pd{\gksnorm}{t}\right)_{2} + \left(\pd{\gksnorm}{t}\right)_{3},
\eeq
with
\begin{gather}
\label{eqn:explicitGKE}
\left(\pd{\gksnorm}{t}\right)_{1} + i \omega_{d,\mbf{k},s}\left(\gksnorm + \frac{Z_s}{\tempnorm} \bess \phinorm \maxsnorm \right) +i \omega_{*,\mbf{k},s} \bess \phinorm  + \mathcal{N}_{\mbf{k},s} = 0, \\
\label{eqn:mirrorGKE}
\left(\pd{\gksnorm}{t}\right)_{2} - \frac{\vths}{\vthr}\musnorm \bhat \cdot \gradnorm \bnorm \pd{\gksnorm}{\vpasnorm} = 0, \\
\label{eqn:streamGKE}
\left(\pd{\gksnorm}{t}\right)_{3} + \frac{\vths}{\vthr}\vpasnorm\bhat\cdot\gradnorm z\left(\pd{\gksnorm}{z}+\frac{Z_s}{\tempnorm}\pd{\bess\phinorm}{z} \maxsnorm \right) = 0.
\end{gather}

Symbolically we can write
\beq
\pd{\mbf{g}}{t} = \mathcal{A}[\mbf{g}]+\left(\mathcal{B} +\mathcal{C}\right)\mbf{g}
\label{eqn:dgdt_operators}
\eeq
with $\mbf{g}$ a vector whose components are the values of $g$ evaluated for the various species and phase space locations, $\mathcal{B}$ and $\mathcal{C}$ are matrices corresponding to the linear operators defined by Eqs.~(\ref{eqn:mirrorGKE}) and~(\ref{eqn:streamGKE}), respectively, and $\mathcal{A}$ is the nonlinear operator defined by Eq.~(\ref{eqn:explicitGKE}).  There is no explicit mention of $\phitb$ in Eq.~(\ref{eqn:dgdt_operators}), as $\phitb$ itself can be expressed via quasineutrality~(\ref{eqn:QNstella}) as an operator acting on $\mbf{g}$.  Discretizing in time and splitting the operators gives
\begin{gather}
\label{eqn:Aoperator}
\mbf{g}^{\nbar} = \mbf{g}^n  + \dt \mathcal{A}[\mbf{g}] \\
\label{eqn:Boperator}
\mbf{g}^{\nbarbar} = \mbf{g}^{\nbar} + \dt \mathcal{B}\mbf{g} \\
\label{eqn:Coperator}
\mbf{g}^{n+1} = \mbf{g}^{\nbarbar} + \dt \mathcal{C} \mbf{g},
\end{gather}
where $n$ indicates the time index and $\dt=\tnorm_{n+1}-\tnorm_n$.  We leave specification of the time discretization of the righthand sides of each of these equations to dedicated subsections below.  The Lie splitting given above is accurate to first order in $\dt$.  Reversing the order of operations -- operating first with $\mathcal{C}$, then $\mathcal{B}$ and finally $\mathcal{A}$ -- in the next time step makes the splitting accurate to second order in $\dt$ when $\mathcal{A}$ is also a linear operator (cf.~\cite{boyd}), albeit with the effective time step doubled in size.  When the nonlinearity is included in $\mathcal{A}$, only first order accuracy is guaranteed.  This `flip-flop' version of Lie splitting is what we employ in \sstella.

\subsection{Explicit time advance for $\mathcal{A}$}\label{s:explicit}

The evolution of $g$ due to magnetic drifts, background gradient drive, and $E\times B$ nonlinear advection, described by Eqs.~(\ref{eqn:explicitGKE}) and~(\ref{eqn:Aoperator}), is treated explicitly in \sstella.  Although not the default option, users may choose to treat the rest of the terms in the gyrokinetic equation, i.e., parallel streaming and acceleration, explicitly as well.  For the explicit time advance algorithm, the user can choose between standard fourth order Runge-Kutta (RK4) and second or third order strong stability preserving  (SSP) Runge-Kutta (RK2 and RK3) schemes~\citep{gottliebSIAM2001}.  While the overall time advance algorithm is limited to second order accuracy in time, the option to treat the explicit terms with a higher order scheme is provided in order to improve their stability properties.  The SSP schemes are constructed so that they retain the stability properties of the forward Euler method and have been optimized so that they allow for the least restrictive CFL condition possible.  We provide details here for the default scheme in \sstella, which is SSP RK3.  Applying SSP RK3 to Eq.~(\ref{eqn:Aoperator}) gives
\beq
\mbf{g}^{\nbar} = \frac{\mbf{g}^n}{3} + \frac{\dt}{2}\mbf{g}_1^n + \frac{\dt}{6}\left(\mbf{g}_2^n + \mbf{g}_3^n\right),
\eeq
with $\mbf{g}_1^n = \mbf{g}^n + \mathcal{A}[\mbf{g}^n]$, $\mbf{g}_2^n=\mbf{g}_1^n+\mathcal{A}[\mbf{g}_1^n]$, and $\mbf{g}_3^n=\mathcal{A}[\mbf{g}_2^n]$.

Our Fourier spectral treatment in $x$ and $y$ eliminates all differential operators in the linear part of $\mathcal{A}$, making it algebraic.  The nonlinear $E\times B$ advection is treated pseudo-spectrally, with de-aliasing achieved by padding the final third of the Fourier coefficients with zeros~\cite{orszagJAS71}.  The use of Fast Fourier Transforms makes each of the $1D$ transforms in $x$ and $y$ computationally efficient ($\mathcal{O}(N\ln N)$ operations, with $N$ the number of padded $k_x$ or $k_y$ coefficients retained).  The explicit time advance is parallelized in \stella so that each $\vpa$, $\mu$ and species can be solved independently, with the only communication occurring at the end of each fractional Runge-Kutta step when the potential $\phitb$ must be updated.

\subsection{Semi-Lagrange treatment of $\mathcal{B}$}

The parallel acceleration described by Eqs.~(\ref{eqn:mirrorGKE}) and~(\ref{eqn:Boperator}) is simply advection in $\vpa$.  The semi-Lagrange approach employed in \stella exploits the fact that this advection has the analytical solution 
\beq
\gksnorm(\vshift,\musnorm,\tnorm+\dt)=\gksnorm(\vpasnorm+\dt(\vths/\vthr)\musnorm\unit{b}\cdot\gradnorm \bnorm,\musnorm,\tnorm),
\label{eqn:mirror_solution}
\eeq
The quantity $\vshift=\vpasnorm+\dt(\vths/\vthr)\musnorm\unit{b}\cdot\gradnorm\bnorm$, does not in general coincide with a grid location in $\vpasnorm$.  We thus approximate the value of $g$ at $\vshift$ by interpolating the values from the four nearest-neighbor grid points, an approach that is accurate to fourth order in $\vpa$ grid spacing, $\dvpa$.  For $\vshift$ falling between grid points $\tld{v}_{\parl,j}$ and $\tld{v}_{\parl,j+1}$, the interpolation formula is
\beq
\tld{g}(\vshift) = \frac{1}{6}\left(c_j c_{j+1} c_{j+2} \tld{g}_{j-1} - 3c_{j-1}c_{j+1}c_{j+2} \tld{g}_{j} + 3c_{j-1} c_j c_{j+2}\tld{g}_{j+1} - c_{j-1}c_j c_{j+1} \tld{g}_{j+2} \right)+ \mathcal{O}(\dvpa^4),
\eeq
with $c_j\defeq(\tld{v}_{\parl,j}-\vshift)/\dvpa$.  At the boundaries in $\vpanorm$, simple linear interpolation is used.  Note that all phase space indices aside from the one corresponding to $\vpa$ have been suppressed for simplicity of notation.  Combining this interpolation formula with the analytical solution~(\ref{eqn:mirror_solution}) and applying it to the split equation~(\ref{eqn:Boperator}) gives
\beq
\tld{g}_{\mbf{k},s,j}^{\nbarbar} = \frac{1}{6}\left(c_j c_{j+1} c_{j+2} \tld{g}_{\mbf{k},s,j-1}^{\nbar} - 3c_{j-1}c_{j+1}c_{j+2} \tld{g}_{\mbf{k},s,j}^{\nbar} + 3c_{j-1} c_j c_{j+2}\tld{g}_{\mbf{k},s,j+1}^{\nbar}- c_{j-1}c_j c_{j+1} \tld{g}_{\mbf{k},s,j+2}^{\nbar} \right).
\eeq
To facilitation this interpolation, a re-mapping is done so that $\tld{g}(\vpanorm)$ is available on all processors; information about $g$ at all other phase space locations can be spread over multiple processors and solved for simultaneously.

Note that the semi-Lagrange approached detailed here places neither an accuracy nor a stability restriction on the time step size: the only error comes from interpolation in $\vpasnorm$, which can be carried out to high order with relatively little numerical expense.

\subsection{Implicit treatment of $\mathcal{C}$}

The dynamics of parallel streaming described by Eqs.~(\ref{eqn:streamGKE}) and~(\ref{eqn:Coperator}) are treated implicitly in \stella following a similar approach to that taken by the local, $\delta f$ gyrokinetic code $\texttt{GS2}$~\cite{kotschCPC95}.  We discretize Eq.~(\ref{eqn:streamGKE}) using variable centering in $z$ and $t$.  For $z$, we use a compact, two-point stencil to facilitate the use of tridiagonal matrix solution methods.  Derivatives in $z$ are given by
\beq
\left(\pd{\tldg}{z}\right)_{i*} = \frac{\tldg_{i+1}-\tldg_i}{\dz},
\label{eqn:dgdz}
\eeq
where the subscripts $i$ and $i+1$ denote evaluation at grid locations $z_i$ and $z_{i+1}$, respectively, and $\dz\defeq z_{i+1}-z_i$.  In Eq.~(\ref{eqn:dgdz}) and for the remainder of this subsection we suppress all phase space indices except those corresponding to $z$ and $t$ to simplify notation.  The subscript $i*$ indicates evaluation at
\beq
z_{i*} \defeq \frac{1\mp\zupwnd}{2} z_i + \frac{1\pm\zupwnd}{2}z_{i+1},
\eeq
with the top (bottom) signs used when the parallel advection speed is positive (negative).  This sign convention will be used for the remainder of this subsection.  The user-specified parameter $\zupwnd$ controls spatial centering: at the extremes, $\zupwnd=0$ corresponds to a centered derivative that is accurate to second order in $\dz$, and $\zupwnd=1$ corresponds to a fully upwinded derivative that is accurate to first order in $\dz$.  All other $z$-dependent quantities are evaluated at $z_{i*}$ using the approximation
\beq
\tldg_{i*} = \frac{1 \mp \zupwnd}{2}\tldg_i +\frac{1 \pm \zupwnd}{2}\tldg_{i+1},
\eeq
which is accurate to second order in $\dz$.

The time discretization is treated in a manner analogous to the $z$ discretization, with implicitness taking the place of upwinding.  The time derivative is given by
\beq
\left(\pd{\tldg}{t}\right)^{n*} = \frac{\tldg^{n+1}-\tldg^{\nbarbar}}{\dt},
\eeq
with the superscript $n*$ indicating evaluation at
\beq
t_{n*} \defeq \frac{1-\tupwnd}{2} t_n+ \frac{1+\tupwnd}{2}t_{n+1}.
\eeq
The user-specified parameter $\tupwnd$ controls temporal centering: at the extremes, $\tupwnd=0$ corresponds to a centered derivative that is accurate to second order in $\dt$, and $\tupwnd=1$ corresponds to a fully implicit treatment accurate to first order in $\dt$.  All other $t$-dependent quantities are evaluated at $t_{n*}$ using the approximation
\beq
\tldg^{n*} = \frac{1 - \tupwnd}{2}\tldg^{\nbarbar} +\frac{1 + \tupwnd}{2}\tldg^{n+1},
\eeq
which is accurate to second order in $\dt$.

Applying the above $z$ and $t$ discretizations to Eq.~(\ref{eqn:streamGKE}) yields
\beq
\begin{split}
\Bigg(&\frac{1\pm\zupwnd}{2}\tldg_{i+1}^{n+1}+\frac{1\mp\zupwnd}{2}\tldg_{i}^{n+1}\Bigg)  +\frac{1+\tupwnd}{2}\frac{\dt}{\dz} \frac{\vths}{\vthr}\vpanorm\left(\bhat\cdot\gradnorm z\right)_{i*}
\left(\tldg_{i+1}^{n+1}-\tldg_{i}^{n+1} 
+\left(J_{0,i+1}\tldphi_{i+1}^{n+1}-J_{0,i}\tldphi_{i}^{n+1}\right)\frac{Z}{\tld{T}} \exp\left(-\tld{v}_{i*}^2\right)\right)\\
&= \left(\frac{1\pm\zupwnd}{2}\tldg_{i+1}^{\nbarbar}+\frac{1\mp\zupwnd}{2}\tldg_{i}^{\nbarbar}\right)  
-\frac{1-\tupwnd}{2}\frac{\dt}{\dz}\frac{\vths}{\vthr}\vpanorm\left(\bhat\cdot\gradnorm z\right)_{i*}\left(\tldg_{i+1}^{\nbarbar}-\tldg_i^{\nbarbar}
+\left(J_{0,i+1}\tldphi_{i+1}^{\nbarbar}-J_{0,i}\tldphi_{i}^{\nbarbar}\right)\frac{Z}{\tld{T}}\exp\left(-\tld{v}_{i*}^2\right)\right)
\label{eqn:stream_discrete}
\end{split}
\eeq
In principle, solving Eq.~(\ref{eqn:stream_discrete}) involves the solution of a linear system that is bidiagonal in the $z$-component of the matrix and is dense in $(\vpa,\mu,s)$ due to the velocity space integral and species sum implicit in $\phitb$.  The inversion of the dense matrix can be avoided by using a Green's function approach~\cite{kotschCPC95}, leaving only a computationally inexpensive bidiagonal matrix solve.

To formulate the Green's function approach, we start by noting that Eq.~({\ref{eqn:stream_discrete}) is linear in $\tldg^{n+1}$.  We can thus express $\tldg^{n+1}$
as the linear combination $\tldg^{n+1}=\tldg_{h}^{n+1}+\tldg_{inh}^{n+1}$, with
\beq
\begin{split}
\Bigg(&\frac{1\pm\zupwnd}{2}\tldg_{inh,i+1}^{n+1}+\frac{1\mp\zupwnd}{2}\tldg_{inh,i}^{n+1}\Bigg)  +\frac{1+\tupwnd}{2}\frac{\dt}{\dz} \frac{\vths}{\vthr}\vpanorm\left(\bhat\cdot\gradnorm z\right)_{i*}
\left(\tldg_{inh,i+1}^{n+1}-\tldg_{inh,i}^{n+1} \right)\\
&= \left(\frac{1\pm\zupwnd}{2}\tldg_{i+1}^{\nbarbar}+\frac{1\mp\zupwnd}{2}\tldg_{i}^{\nbarbar}\right)  
-\frac{1-\tupwnd}{2}\frac{\dt}{\dz}\frac{\vths}{\vthr}\vpanorm\left(\bhat\cdot\gradnorm z\right)_{i*}\left(\tldg_{i+1}^{\nbarbar}-\tldg_i^{\nbarbar}
+\left(J_{0,i+1}\tldphi_{i+1}^{\nbarbar}-J_{0,i}\tldphi_{i}^{\nbarbar}\right)\frac{Z}{\tld{T}}\exp\left(-\tld{v}_{i*}^2\right)\right)
\label{eqn:stream_inh}
\end{split}
\eeq
and
\beq
\Bigg(\frac{1\pm\zupwnd}{2}\tldg_{h,i+1}^{n+1}+\frac{1\mp\zupwnd}{2}\tldg_{h,i}^{n+1}\Bigg)  +\frac{1+\tupwnd}{2}\frac{\dt}{\dz} \frac{\vths}{\vthr}\vpanorm\left(\bhat\cdot\gradnorm z\right)_{i*}
\left(\tldg_{h,i+1}^{n+1}-\tldg_{h,i}^{n+1} 
+\left(J_{0,i+1}\tldphi_{i+1}^{n+1}-J_{0,i}\tldphi_{i}^{n+1}\right)\frac{Z}{\tld{T}} \exp\left(-\tld{v}_{i*}^2\right)\right)
= 0.
\label{eqn:stream_h}
\eeq
The `twist-and-shift' boundary condition described in Sec.~\ref{s:realgrids} is applied at the end of each $2\pi$ segment in $z$.  This boundary condition couples multiple $2\pi$ segments in $z$ with different $k_x$ values, leading to an extended $z$ domain with $N_{z+}$ points, where $N_{z+}=N_z \times N_{\textnormal{seg}}$ and $N_{\textnormal{seg}}$ is the number of connected segments.

From quasineutrality, we have
\beq
\tldphi_i^{n+1} = \left(\sum_s \left(1-\Gamma_{0,i,s}\right)\frac{Z_s^2}{\tempnorm} \nnorm \right)^{-1}\sum_s Z_s\nnorm \frac{2\bnorm}{\pi^{1/2}} \int_{-\infty}^{\infty}d\vpasnorm\int_0^{\infty}d\musnorm J_{0,i,s} \tldg_{i,s}^{n+1}.
\label{eqn:QNphi}
\eeq

We get the Green's function for $\tldg_{h}^{n+1}$ by supplying a unit impulse to $\tldphi$ for each $z$ location in the extended $z$ domain and solving Eq.~(\ref{eqn:stream_h}) for the response $\tldg_h^{n+1}$.  Following this approach we have
\beq
\tldg_i^{n+1} = \sum_{p=1}^{N_{z+}}\frac{\delta \tldg_{h,i}}{\delta \tldphi_p}\tldphi_p^{n+1} + \tldg_{inh,i}^{n+1},
\eeq
where $N_{z+}$ is the number of grid points in the extended $z$ domain, and $\delta \tldg_{h,i}/\delta \tldphi_p$ is the response of $\tldg_h^{n+1}$ at grid location $z_i$ to a unit perturbation in $\tldphi^{n+1}$ at grid location $z_p$.  Substituting this form for $\tldg_i^{n+1}$ into Eq.~(\ref{eqn:QNphi}) yields an implicit equation for the vector $\bm{\phitb}^{n+1}$ whose $i^{th}$ component is $\tldphi_i^{n+1}$:
\beq
\left(I - Q \sum_{p=1}^{N_{z+}}\frac{\bm{\delta g}^{n+1}}{\bm{\delta \phitb}^{n+1}}\right)\bm{\phitb}^{n+1} = \bm{\phitb}_{inh}^{n+1},
\label{eqn:stream_phi}
\eeq
where $(\bm{\delta g}/\bm{\delta \phitb})_{ip}=\delta \tldg_{h,i}/\delta \tldphi_p$, $I$ is the $N_{z+}\times N_{z+}$ identity matrix,
\beq
Q=\left(\sum_s \left(1-\Gamma_{0s}\right)\frac{Z_s^2 e}{T_s} n_s \right)^{-1}\sum_s Z_s \int d^3v J_{0,s}
\eeq
is the velocity-space operator appearing in quasineutrality and $(\bm{\phitb}_{inh})_i = Q \tldg_{inh,i}$.

Thus \stella first solves Eq.~(\ref{eqn:stream_inh}) for $\tldg_{inh}^{n+1}$ and uses it in Eq.~(\ref{eqn:stream_phi}) to obtain $\tldphi^{n+1}$ via $LU$ decomposition and back-substitution.  Finally, the updated distribution function $\tldg^{n+1}$ is calculated via Eq.~(\ref{eqn:stream_discrete}).  The layout of the data for this implicit solve is the same as for the explicit advance described in Subsection~\ref{s:explicit}: information for each $k_x$, $k_y$, and $z$ are available to all processors, while $g$ evaluated at each $\vpa$, $\mu$, and species can be solved for simultaneously.

\subsubsection{Zonal modes}

The $k_y=0$ modes, often referred to as a zonal modes, must be treated specially, as they are periodic in $z$.  To enforce periodicity, we solve the gyrokinetic equation twice each time it is required: once with a zero incoming boundary condition in $z$; and once with a unity incoming boundary condition in $z$, but with no terms involving $\tldg^{\nbarbar}$ or $\tldphi^{\nbarbar}$.  We denote the former solution as $\tldg_{PI}$ and the latter as $\tldg_{CF}$.  Noting that any linear combination of $\tldg_{PI}$ and $\tldg_{CF}$ is also a solution for $\tldg$ and enforcing periodicity, we have
\beq
\tldg_1^{n+1} = \tldg_{PI,1} + d \tldg_{CF,1} = d = \tldg_{PI,N_z} + d \tldg_{CF,N_z}.
\eeq
Solving for $d$ and substituting into the above linear combination gives the solution for zonal modes:
\beq
\tldg_i^{n+1} = \tldg_{PI,i} + \frac{\tldg_{PI,N_z}}{1-\tldg_{CF,N_z}} \tldg_{CF,i}.
\eeq


\section{Numerical tests}\label{s:benchmarks}

In this section we provide simulation data to illustrate the accuracy and efficiency of \sstella.  Throughout, we verify the \stella simulations by comparing with the widely-benchmarked, $\delta f$-gyrokinetic code \gs~\cite{kotschCPC95,dorlandPRL00}.  We focus on two magnetic field configurations for our comparisons: the first is the so-called `Cyclone Base Case' (CBC), a widely-used benchmark case~\cite{dimitsPoP00} in the magnetic confinement fusion community that has an axisymmetric magnetic field with concentric circular flux surfaces; the second is design LI383 for the National Compact Stellarator Experiment (NCSX), a case which has also been used for benchmarking within the stellarator community~\cite{baumgaertelPoP11}.

All simulations for the CBC used the Miller local equilibrium~\cite{millerPoP98} option to specify geometric coefficients, while the NCSX simulations used data from the \texttt{VMEC}-generated equilibrium for LI383.  Tables~\ref{t:cbc} and~\ref{t:ncsx} provide the relevant \stella input parameters for these cases.

\begin{table}
\centering
\begin{tabular}{ |p{2.4cm}|p{1.6cm}| }
 \hline
 \multicolumn{2}{|c|}{CBC input parameters} \\
 \hline
 Input variable & Input value \\
 \hline
 $\tld{r}=r/a$ & 0.5 \\
 $\tld{R}=R_0/a$ & 2.77778 \\
 $d\tld{R}/d\tld{r}$ & 0 \\
 $q$ & 1.4 \\
 $\hat{s}=d\ln q/d\ln r$ & 0.796 \\
 $\kappa$ & 1.0 \\
 $d\kappa/d\tld{r}$ & 0.0 \\
 $\delta$ & 0.0 \\
 $d\delta/d\tld{r}$ & 0.0 \\
 $B_r$ & $B_{\zeta}(R_0)$ \\
 $n_i/n_e$ & 1.0\\
 $T_i/T_e$ & 1.0\\
 $m_i/m_e$ & 3672\\ $d\ln n_i/d\tld{r}$ & -0.8 \\
 $d\ln n_e/d\tld{r}$ & -0.8 \\
 $d\ln T_i/d\tld{r}$ & -2.49 \\
 $d\ln T_e/d\tld{r}$ & -2.49 \\
 \hline
\end{tabular}
\caption{List of \stella input parameters for the CBC simulations}
\label{t:cbc}
\end{table}

\begin{table}
\centering
\begin{tabular}{ |p{3cm}|p{1.6cm}| }
 \hline
 \multicolumn{2}{|c|}{NCSX design LI383 input parameters} \\
 \hline
 Input variable & Input value \\
 \hline
 $\tld{\psi}=\psi_t/\psi_{t,LCFS}$ & 0.635 \\
 $\alpha$ & 0 \\
 \# of field periods & 3 \\
 $n_i/n_e$ & 1.0\\
 $T_i/T_e$ & 1.0\\
 $m_i/m_e$ & 3672\\
 $d\ln n_i/dx$ & 0.0 \\
 $d\ln T_i/dx$ & 4.0 \\
 \hline
\end{tabular}
\caption{List of \stella input parameters for the NCSX simulations}
\label{t:ncsx}
\end{table}

\subsection{Linear simulation results for the CBC}\label{s:lincbc}

We start by comparing linear growth rates, frequencies, and mode structures obtained from \stella and \gs simulations for the CBC.  
Unless stated otherwise, all linear CBC simulation results shown here were obtained with the following resolution: Both \stella and \gs used $N_z=25$ and three $2\pi$ segments in an extended ballooning domain.  Additionally, $\texttt{stella}$ used $N_{\vpa}=48$, $N_{\mu}=12$, and $\tld{v}_{\parl,c}=\tld{v}_{\perp,c}=3$, while \gs used 33 pitch angles (20 in the untrapped region of phase space and 13 in the trapped region) and 16-32 energy grid points.  The \stella cell-centering parameters in $z$ and $t$ were set to $u_z=u_t=0.02$.  A description of the \gs velocity space treatment can be found in Ref.~\cite{barnesPoP10a}.  

\begin{figure}[htpb]
  \begin{center}
    \includegraphics*[width=0.49\textwidth]{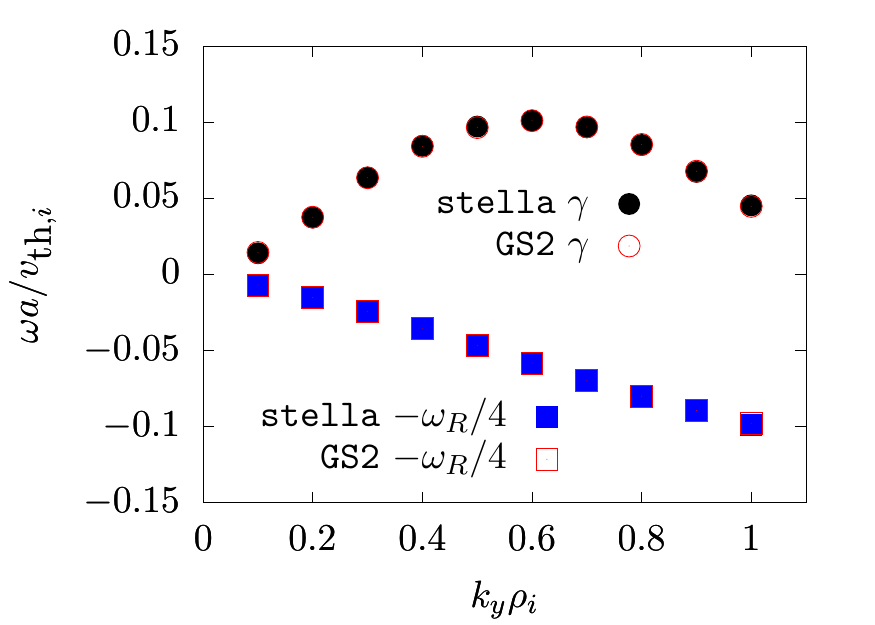}
        \includegraphics*[width=0.49\textwidth]{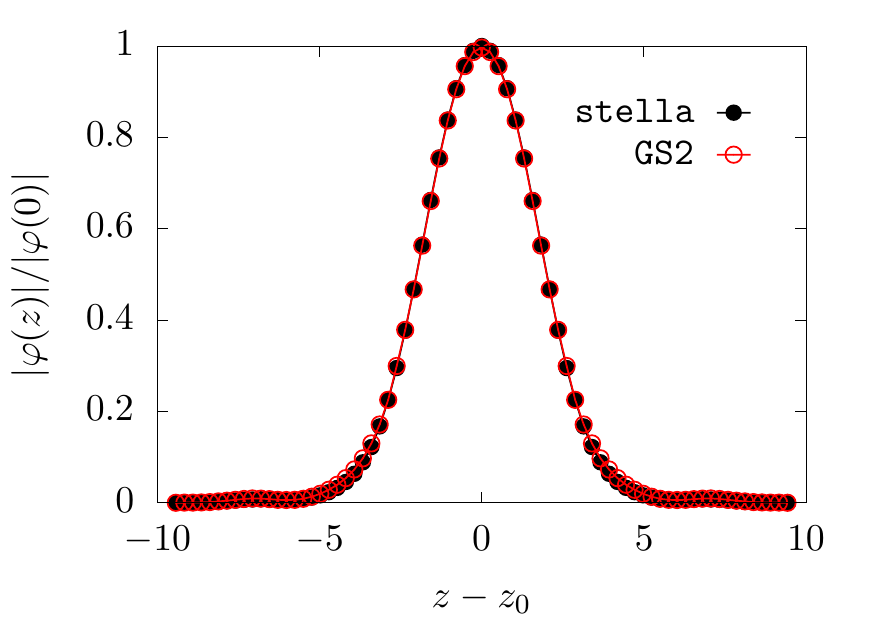}
  \end{center}
  \caption{(Left): Normalized real (\textit{squares}) and imaginary (\textit{circles}) components of mode frequency $\omega$ as a function of the normalized binormal wavenumber $k_y$ for $\texttt{stella}$ (\textit{filled blue/black}) and $\texttt{GS2}$ (\textit{open red}). (Right): Electrostatic potential for the fastest growing mode ($k_y\rho_i=0.6$) as a function of the extended parallel coordinate $z-z_0=\theta-\theta_0$ for $\texttt{GS2}$ (\textit{open red circles}) and $\texttt{stella}$ (\textit{filled black circles}).
These data correspond to CBC parameters with a modified Boltzmann response for electrons.}
  \label{fig:cbc_ae_om}
\end{figure}

First, we compare the two codes for the case of a modified Boltzmann response for the electrons; i.e.,  $\delta n_e = en_e(\phitb-\overline{\phitb})/T_e$, with the overline denoting an average along the magnetic field.  The results are given in Fig.~\ref{fig:cbc_ae_om}.  All quantities agree to within a few percent across the entire range of unstable $k_y$ values.  We compare the same quantities with kinetic electrons at both ion and electron scales in Figs.~\ref{fig:cbc_keion_om} and~\ref{fig:cbc_keelectron_om}.  In both cases, there is again excellent agreement between the data from \stella and \gs.

\begin{figure}[htpb]
  \begin{center}
    \includegraphics*[width=0.49\textwidth]{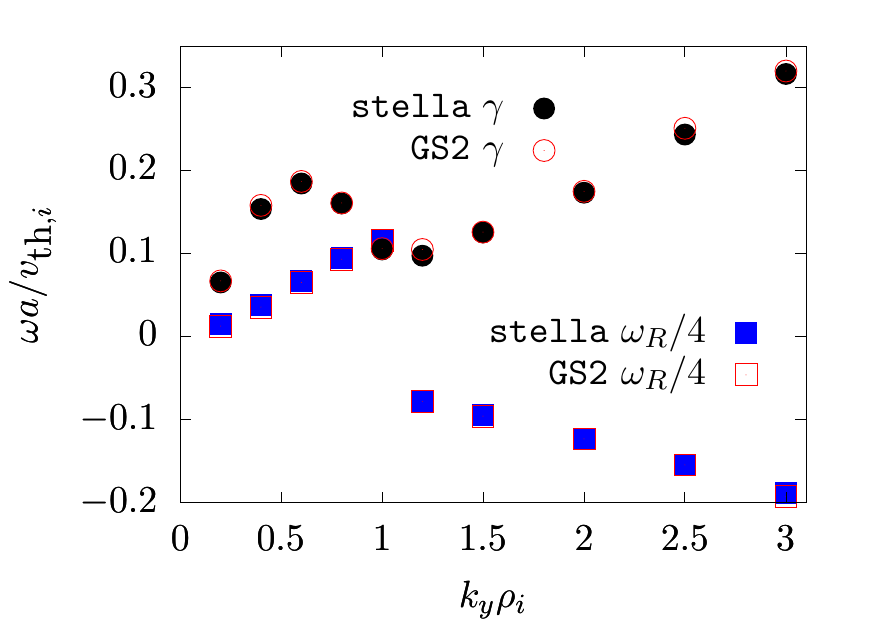}
    \includegraphics*[width=0.49\textwidth]{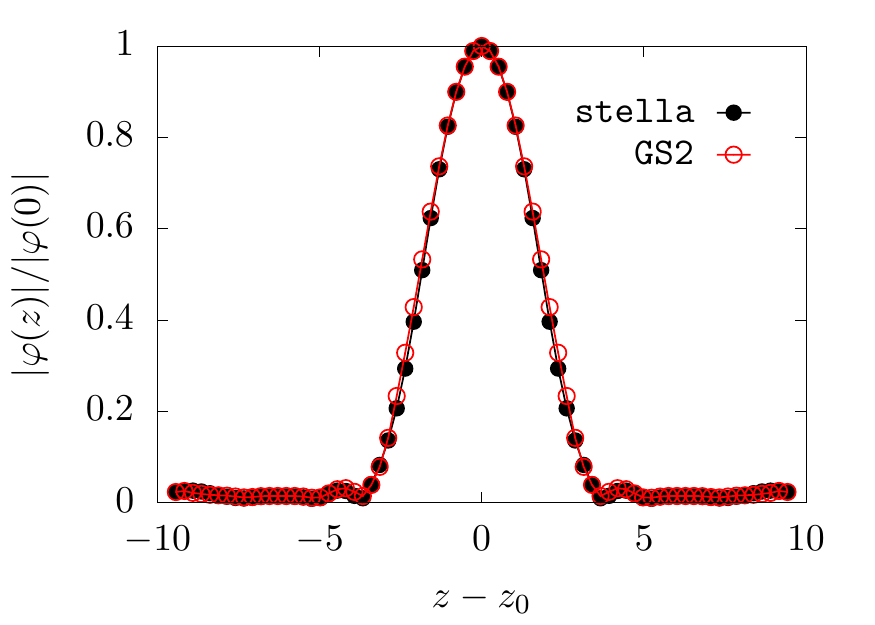}
  \end{center}
  \caption{(Left): Normalized real (\textit{squares}) and imaginary (\textit{circles}) components of mode frequency $\omega$ as a function of the normalized binormal wavenumber $k_y$ for $\texttt{stella}$ (\textit{filled blue/black}) and $\texttt{GS2}$ (\textit{open red}).  (Right): Electrostatic potential for the fastest growing mode ($k_y\rho_i=0.6$) as a function of the extended parallel coordinate $z-z_0=\theta-\theta_0$ for $\texttt{GS2}$ (\textit{open red circles}) and $\texttt{stella}$ (\textit{filled black circles}).  These data correspond to CBC parameters with kinetic electrons.}
  \label{fig:cbc_keion_om}
\end{figure}

\begin{figure}[htpb]
  \begin{center}
    \includegraphics*[width=0.49\textwidth]{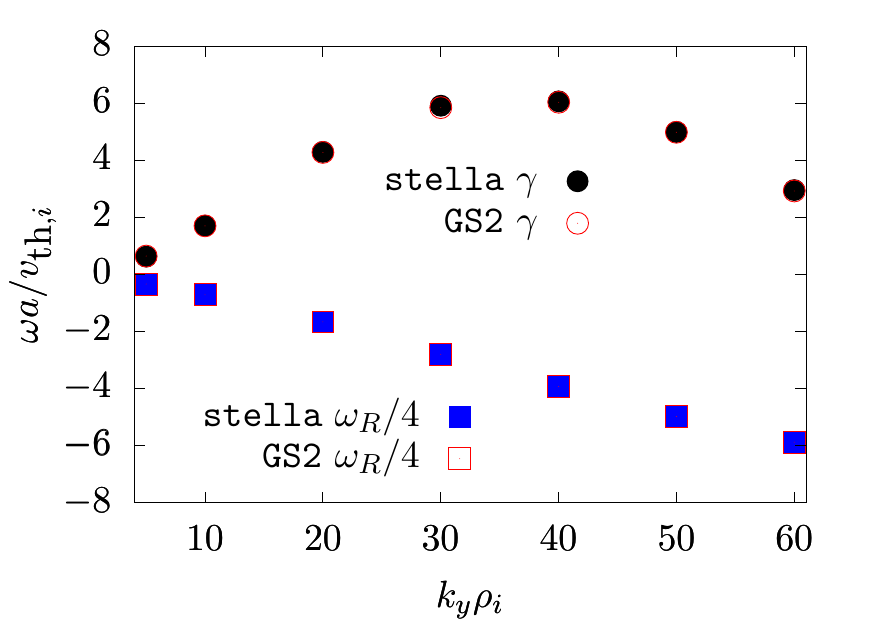}
    \includegraphics*[width=0.49\textwidth]{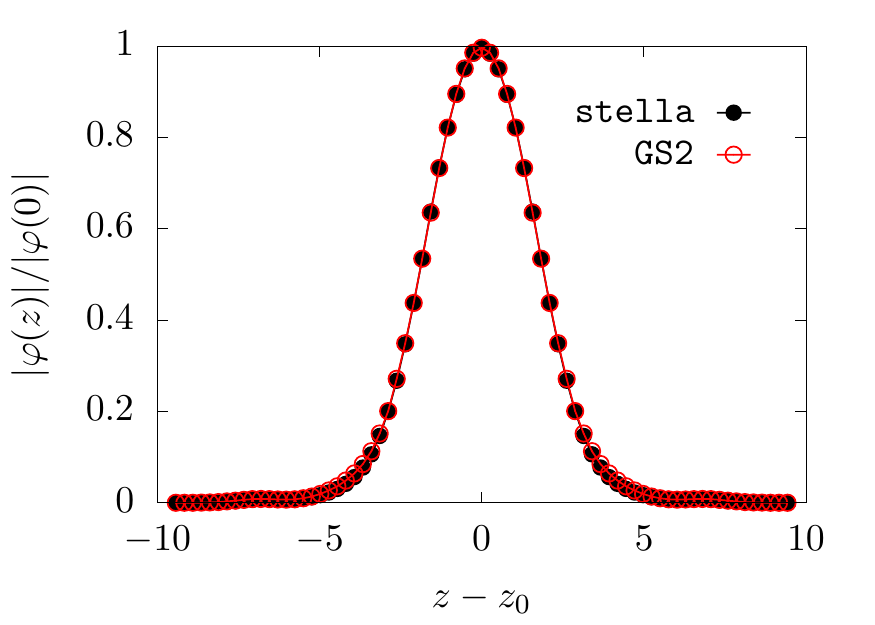}
  \end{center}
  \caption{(Left): Normalized real (\textit{squares}) and imaginary (\textit{circles}) components of mode frequency $\omega$ as a function of the normalized binormal wavenumber $k_y$ for $\texttt{stella}$ (\textit{filled blue/black}) and $\texttt{GS2}$ (\textit{open red}).  (Right): Electrostatic potential for the fastest growing mode at electron scales ($k_y\rho_i=40$) as a function of the extended parallel coordinate $z-z_0=\theta-\theta_0$ for $\texttt{GS2}$ (\textit{open red circles}) and $\texttt{stella}$ (\textit{filled black circles}).  These data correspond to CBC parameters with kinetic electrons.}
  \label{fig:cbc_keelectron_om}
\end{figure}

As noted in Section~\ref{s:algorithm}, a guiding principle for the \stella algorithm was the desirability of an implicit treatment for parallel streaming and acceleration when including kinetic electrons.  To demonstrate the utility of the mixed implicit-explicit (IMEX) treatment detailed in Section~\ref{s:algorithm}, we determined the maximum stable time step for both the fully explicit and the IMEX versions of \sstella.  The results are shown in Fig.~\ref{fig:dt_scaling}.  As mentioned in Section~\ref{s:algorithm}, the fully explicit approach has a much more restrictive CFL condition than the IMEX approach:  For $k_y\rho_i \sim 1$, the maximum stable time step for the explicit scheme is $\sim 100$ times smaller than for the IMEX scheme, and this gap widens at longer wavelengths.  

The severe time step constraint for $k_y\rho_i \ll 1$ is due to the rapid response of the electric field to small charge imbalances at long wavelength.  This response leads to a high frequency mode in the plasma known as the shear-Alfv\'en wave~\cite{wwleeJCP87}, which has a frequency proportional to $1/k_{\perp}$.  However, an even more restrictive constraint on the time step that scales as $1/k_{\perp}^2$ appears if one does not use centered differences when discretizing the $z$ derivative of $\phitb$~\cite{candyJCP03}.  A brief calculation deriving these time step constraints for the simplified case of an un-sheared, homogenous plasma slab is given in~\ref{s:appendix}.


At short wavelengths, the explicit treatment of advection by magnetic drifts -- with an advection speed proportional to the wavenumber -- provides a CFL time step that scales as $1/k_y$.  This is evident for $k_y\rho_i \gtrsim 0.4$ in Fig.~\ref{fig:dt_scaling}.  The explicit treatment of the magnetic drifts is also responsible for the CFL time step at long wavelengths in the IMEX approach.  A calculation similar to that given in~\ref{s:appendix} shows that the terms containing a product of the magnetic drifts and the electrostatic potential scale inversely with $k_y$, leading to a CFL time step at long wavelengths that scales as $k_y$.  This scaling is evident for $k_y\rho_i \lesssim 0.3$ in Fig.~\ref{fig:dt_scaling}.

Finally, to demonstrate the utility of the `flip-flop' version of Lie operator spitting discussed in Sec.~\ref{s:algorithm}, we show in Fig.~\ref{fig:dt_scaling} the convergence of the computed complex frequency $\omega$ with decreasing time step $\Delta t$ for a case with Boltzmann electrons.  Here, the numerical resolution used is $N_{\vpa}=64$, $N_{\mu}=6$ and $N_z=9$, with one $2\pi$ segment along $z$.  Denoting $\omega_0$ as the value for $\omega$ at very small $\Delta t$ (a factor of 3 below those shown), we see that the error $|\omega-\omega_0|$ scales as $(\Delta t)^2$ for the `flip-flop' scheme and as approximately $(\Delta t)^{5/4}$ for regular Lie splitting.

\begin{figure}[htpb]
  \begin{center}
  \includegraphics*[width=0.49\textwidth]{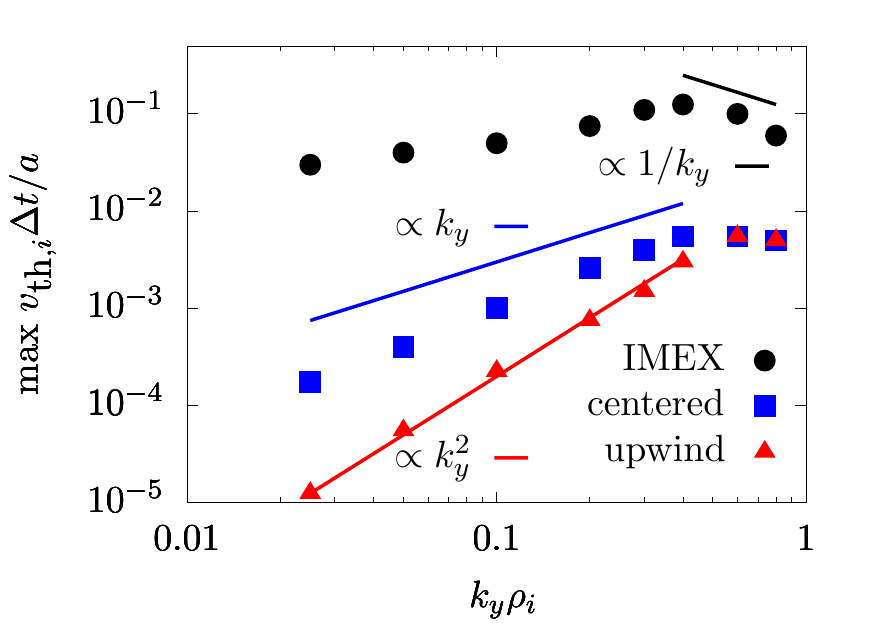}
    \includegraphics*[width=0.49\textwidth]{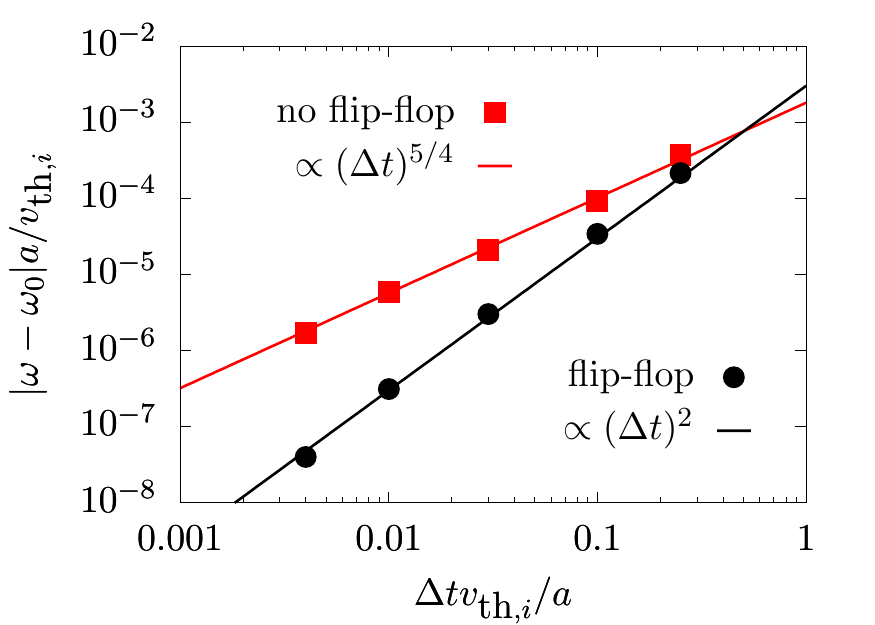}
  \end{center}
  \caption{(Left): Maximum stable time step as a function of $k_y\rho_i$ for the fully explicit scheme with upwind (\textit{red triangles}) and centered (\textit{blue squares}) differencing of $\partial \phitb/\partial z$ and for the mixed implicit-explicit (\textit{black circles}) time advance schemes.  Also shown are the expected scalings of the maximum stable time step with $k_y\rho_i$ (see main text for an explanation of these scalings).  These correspond to CBC parameters with kinetic electrons.  (Right): Modulus of the error in the complex frequency $\omega$ as a function of time step size $\Delta t$ for the time advance scheme with Lie operator splitting (red squares) and with the `flip-flop' variant of Lie splitting (black circles).  The reference complex frequency $\omega_0$ is obtained from a simulation with $\Delta t v_{\textnormal{th},i}/a=0.0015$.  All simulations used CBC parameters with a modified Boltzmann response for electrons.}
  \label{fig:dt_scaling}
\end{figure}



\subsection{Nonlinear simulation results for the CBC}\label{s:nlcbc}

We next compare turbulent heat fluxes from nonlinear simulations with CBC parameters for \stella and \gs.  Both \stella and \gs simulations used $N_z=32$, $N_{k_y}=22$, and $N_{k_x}=128$, with a box size in $x$ and $y$ of approximately $126\rho_i$.  Note that the effective value for $N_{k_y}$ should be doubled, as both \stella and \gs use the reality condition to limit the simulated $\mbf{k}$-domain so that $k_y\geq 0$.  The velocity space resolution was the same for both codes as the linear case, with the exception that \stella used $N_{\mu}=16$.  A small amount of hyper-viscosity was employed in all simulations to avoid spectral pile-up and was treated implicitly using the same `flip-flop' operator splitting employed for the parallel acceleration and streaming.  The form for hyper-viscosity currently used in \stella is
\beq
\pd{g}{t} = -D \frac{k_{\perp}^4}{k_{\perp,max}^4} g,
\eeq
with $D=0.05$ for the simulations reported here.

The turbulent heat fluxes for simulations with kinetic electrons and with a modified Boltzmann response for electrons are given in Fig.~\ref{fig:cbc_ae_nl}.  There is remarkable agreement between the \stella and \gs heat fluxes for ions in both cases and for electrons in the case where they are treated kinetically.

\begin{figure}[htpb]
  \begin{center}
    \includegraphics*[width=0.49\textwidth]{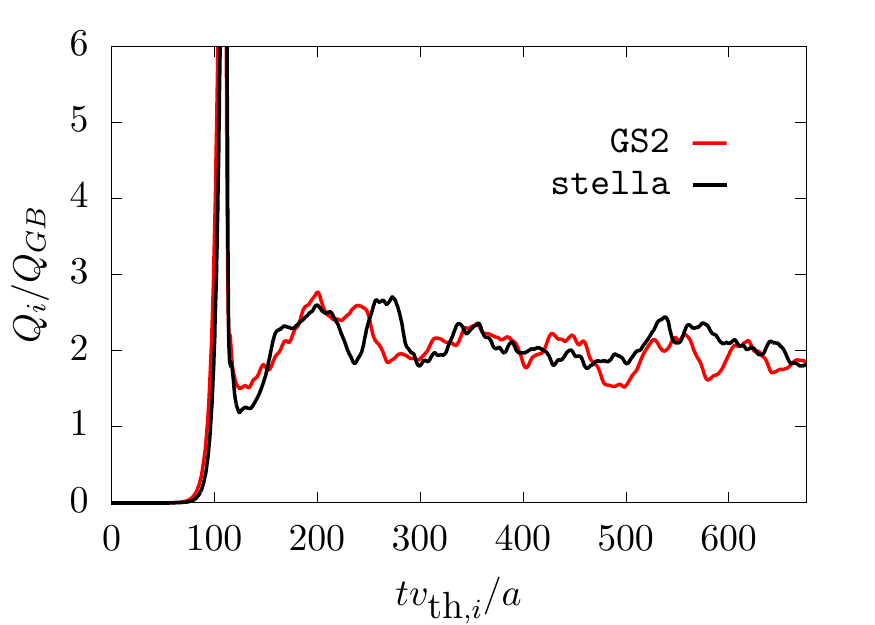}
\includegraphics*[width=0.49\textwidth]{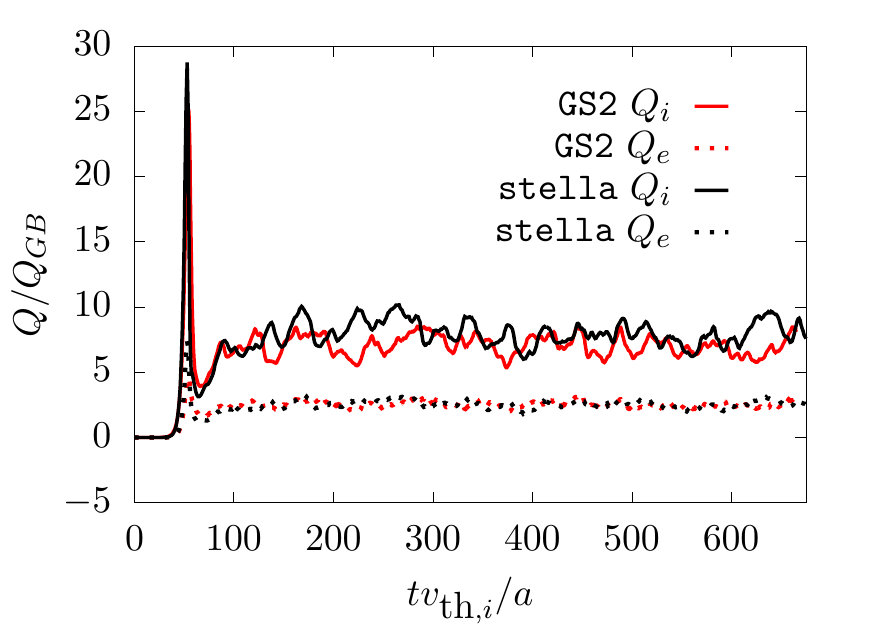}
  \end{center}
  \caption{Time traces of the gyro-Bohm-normalized heat flux ($Q_{GB}=n_iT_i\vthi \rhoi^2/a^2$) for \stella (\textit{black line}) and \texttt{GS2} (\textit{red line}).  Simulations obtained using CBC parameters. (Left): Electrons have a modified Boltzmann response. (Right):  Electrons are treated kinetically.  Solid lines denote ion fluxes, and dashed line denote electron fluxes.}
  \label{fig:cbc_ae_nl}
\end{figure}


\subsection{Linear simulation results for NCSX design LI383}\label{s:ncsx}

For the NCSX linear benchmarks, a modified Boltzmann response was enforced for the electrons.  The resolution required to achieve converged results at all $k_x$ and $k_y$ values was higher than for the CBC simulations in both \stella and \gs.  Both codes included all three NCSX field periods (corresponding to a $\zeta$ domain of [-15.653,15.653]).  The \stella simulations used $N_z=1025$, $N_{\vpa}=128$, and $N_{\mu}=24$, while the \gs simulations used $N_z=455$, 24 energy grid points, and 50 (60) pitch-angles for the $k_y$ ($k_x$) scan.  These extreme resolutions were not necessary in most cases, but were used to ensure agreement in the most challenging ones.

The variation in magnetic field strength with toroidal angle $\zeta$ is given in Fig.~\ref{fig:ncsx_lin_kx0p64_eigenfunc}, along with an example of the mode structure along $z$ for $(k_x=0.636,k_y=1.414)$.  One can see that there is a significant amount of structure, which is what necessitates the higher resolution in $z$.  Plots of the linear growth rates and real frequencies for both $k_x$ and $k_y$ scans are given in Fig.~\ref{fig:ncsx_lin_kxscan}.  

\begin{figure}[htpb]
  \begin{center}
    \includegraphics*[width=0.49\textwidth]{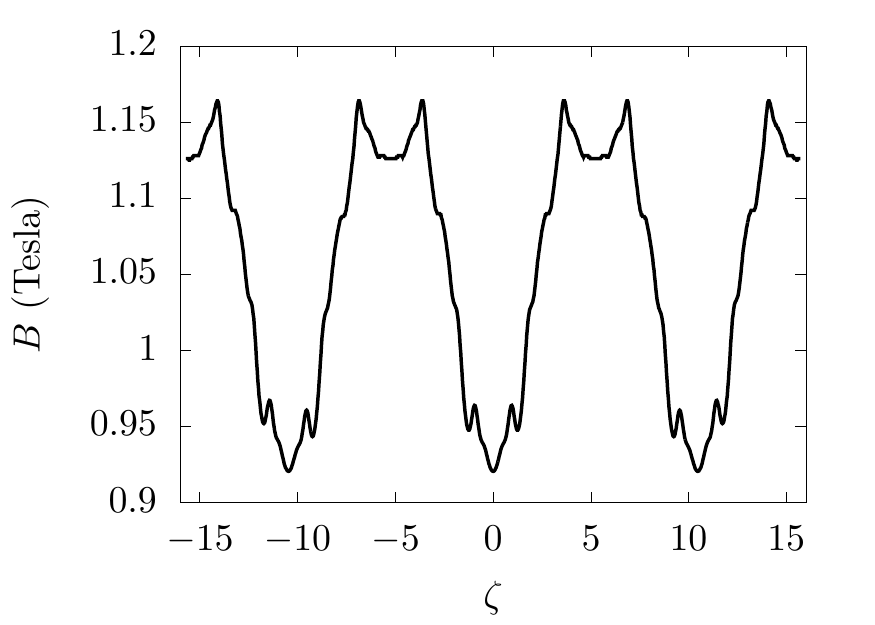}
    \includegraphics*[width=0.49\textwidth]{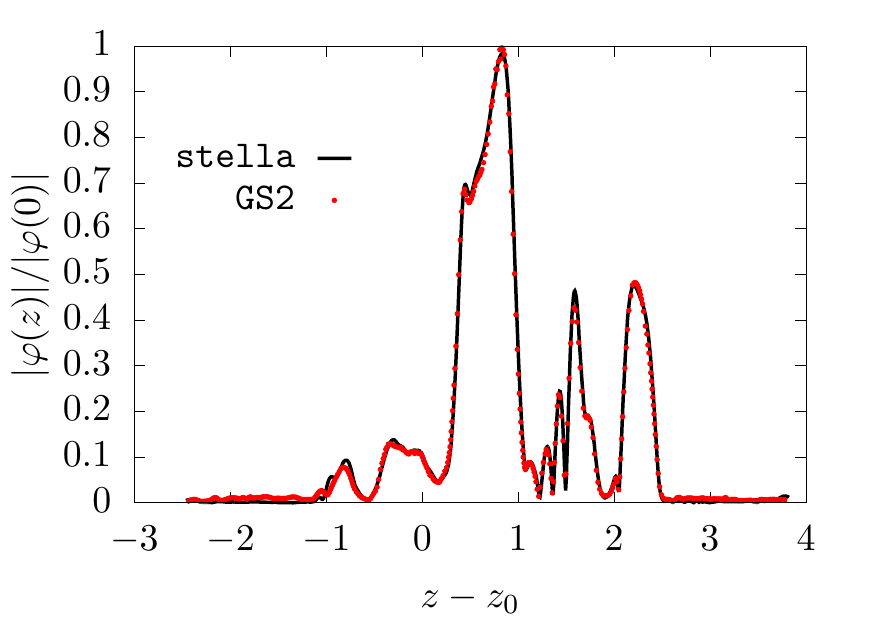}
  \end{center}
  \caption{(Left): Magnetic field strength vs $\zeta$ for the simulated NCSX equilibrium (LI383).  (Right): Normalized modulus of $\phitb$ vs $z-z_0$ for $(k_x\rho_i=0.64,k_y\rho_i=1.414)$ obtained by \stella (\textit{black line}) and \gs (\textit{red circles}).}
  \label{fig:ncsx_lin_kx0p64_eigenfunc}
\end{figure}

\begin{figure}[htpb]
  \begin{center}
    \includegraphics*[width=0.49\textwidth]{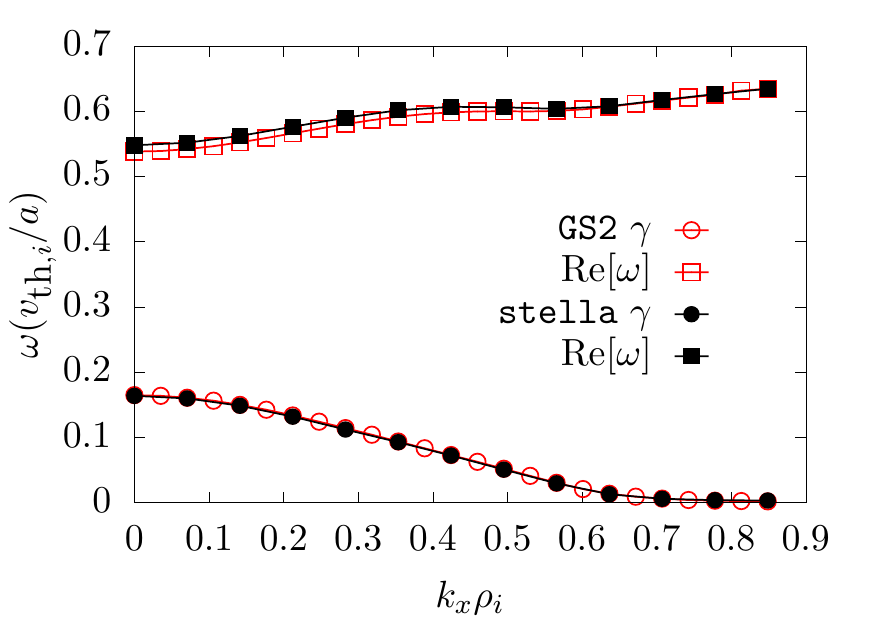}
        \includegraphics*[width=0.49\textwidth]{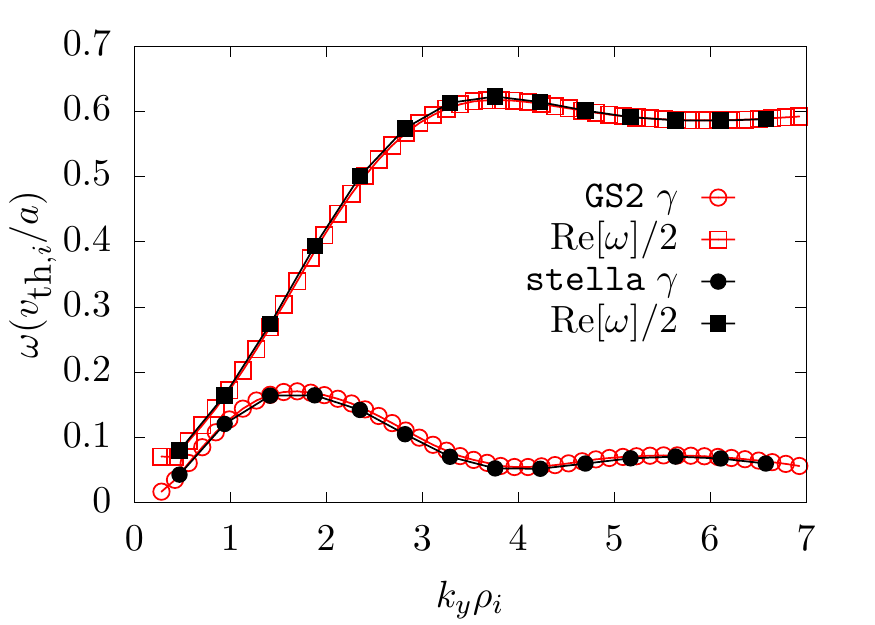}
  \end{center}
  \caption{Real frequency and growth rate spectra for scaled NCSX equilibrium LI383 for $k_y\rho_i=1.414$ (left) and $k_x\rho_i=0$ (right).}
  \label{fig:ncsx_lin_kxscan}
\end{figure}




\section{Summary and future work}\label{s:futurework}

The algorithm for \stella presented here enables fast, accurate evaluation of the gyrokinetic equation~(\ref{eqn:gke}) subject to the quasineutrality constraint~(\ref{eqn:QN}).  Its use of a mixed implicit-explicit (IMEX) algorithm greatly reduces the CFL time step constraint, especially at long wavelengths.  The implicit part of the solve is facilitated by operator splitting, which allows for a flexible treatment of the various different physics effects appearing in the gyrokinetic equation.  The code has been benchmarked both linearly (Sec.~\ref{s:lincbc}) and nonlinearly (Sec.~\ref{s:nlcbc}) -- with and without retention of kinetic electron dynamics -- for axisymmetric magnetic field configurations.  It has also been benchmarked linearly with a modified Boltzmann electron response in a non-axisymmetric magnetic field configuration (Sec.~\ref{s:ncsx}).  As such, there are a wide range of problems to which it can be immediately and usefully applied.

However, there are a few obvious ways in which \stella could be improved that are under development.  The code does not currently include the effect of Coulomb collisions or of magnetic fluctuations, both of which should be straightforward extensions to \sstella.  Also, as pointed out in Sec.~\ref{s:intro}, one of the motivations for developing \stella was to study turbulence that is non-local in the bi-normal $\alpha$ coordinate.  This can be achieved by abandoning the flux tube approach in favor of an annulus that encompasses the full flux surface of interest.  Although this has not been done in the current version of \sstella, the IMEX algorithm with operator splitting was devised with full flux surface simulations in mind.  Implementation of this full flux surface option is in progress and will be addressed in detail in a future publication.

\section*{Acknowledgements}
The authors would like to thank W. Dorland for useful discussions.  The authors acknowledge the use of ARCHER through the Plasma HEC Consortium EPSRC grant number EP/L000237/1 under project e281-gs2 and the use of the EUROfusion High Performance Computer (Marconi-Fusion) under project MULTEI.

\appendix

\section{Long wavelength numerical instability}\label{s:appendix}

Here we address the issue of numerical stability of the gyrokinetic-Poisson system of equations~(\ref{eqn:gkstella}) and~(\ref{eqn:QNstella}) at long wavelengths.  It has been shown previously
that a high frequency mode termed the electrostatic shear-Alfv\'en wave is supported by the plasma within the electrostatic approximation and that this leads to a CFL time step that scales as $k_{\parl}/k_{\perp}$~\cite{wwleeJCP87}.  It has also been noted that one must take care when discretizing the $z$ derivative appearing in the gyrokinetic equation~(\ref{eqn:gkstella}) in order to avoid a numerical instability that scales inversely with $k_{\perp}^2$~\cite{candyJCP03}.  Here we provide a brief calculation illustrating the origin of this numerical instability and reversion to the CFL constraint of the electrostatic Alfv\'en wave in the appropriate limit.

To simplify our analysis, we consider an un-sheared, homogeneous plasma slab.  For such a system the perpendicular speed (or, equivalently, the magnetic moment) appears only as a parameter in the gyrokinetic equation and can thus be averaged away.  The resulting system of equations is
\beq
\pd{G_{\mbf{k},s}}{t} +\vpa\left(\pd{G_{\mbf{k},s}}{z}+\frac{Z_s e n_s}{T_s}\Gamma_0(b_{\mbf{k},s})\pd{\phik}{z} \frac{e^{-\vpa^2/\vths^2}}{\sqrt{\pi}\vths} \right)=0,
\label{eqn:slabGK}
\eeq
\beq
\sum_s Z_s e \left( \int_{-\infty}^{\infty} d\vpa\ G_{\mbf{k},s} + \frac{Z_sen_s}{T_s}\left(\Gamma_0(b_{\mbf{k},s})-1\right)\phik\right) = 0,
\label{eqn:slabQN}
\eeq
with $G_{\mbf{k},s}(\vpa,z,t)\defeq 2\pi\int_0^{\infty}d\vpe \vpe \bess \gks$.  We discretize~(\ref{eqn:slabGK}) and~(\ref{eqn:slabQN}) using a simple, first-order upwind scheme for $G$ and a more general scheme that combines upwind and centered differences for $\phitb$; i.e., 
\beq
\pd{G_j}{t} + \frac{\vpa}{\dz} \left( \sigma\left(G_{j}-G_{j-\sigma}\right)
+ \frac{Z e n}{T} \Gamma_0(b) \left(\alpha \sigma \left(\phitb_{j}-\phitb_{j-\sigma}\right)+\frac{(1-\alpha)}{2}\left(\phitb_{j+1}-\phitb_{j-1}\right)\right)\frac{e^{-\vpa^2/\vth^2}}{\sqrt{\pi}\vth}\right) = 0,
\eeq
where $\sigma\defeq \vpa/|\vpa|$, $\alpha \in [0,1]$ controls the balance of upwind and centered differences in the evaluation of $\partial \phitb/\partial z$, and the subscript $j$ denotes evaluation at $z_j$.  Note that we have suppressed species and wavenumber subscripts to simplify notation.

Assuming solutions for $G$ of the form $G_j=\hat{G}(\vpa)\exp\left(ik_{\parl}z_j -i\omega t\right)$ then leads to the dispersion relation
\beq
\sum_s \frac{Z_s^2 n_s}{T_s} \left(1-\Gamma_{0}(b_{\mbf{k},s})\right) = \sum_s \frac{Z_s^2 n_s}{T_s}\frac{\Gamma_0(b_{\mbf{k},s})}{\sqrt{\pi}}\int_{-\infty}^{\infty} du \left(\frac{\alpha S_1 + (1-\alpha)S_2}{\zeta_s - u S_1}\right)ue^{-u^2},
\label{eqn:dispersion}
\eeq
where $u\defeq \vpa/\vths$, $\zeta_s\defeq \omega/k_{\parl}\vths$, $S_1\defeq \textnormal{sinc}(k_{\parl}\dz/2)\exp(-i\sigma k_{\parl}\dz/2)$, and $S_2\defeq \textnormal{sinc}(k_{\parl}\dz)$.  Anticipating the existence of high frequency modes at long wavelength, we expand~(\ref{eqn:dispersion}) for $\zeta_s \gg 1$ and $k_{\perp}\rho_s \ll 1$ to obtain
\beq
\sum_s \frac{Z_s^2 n_s}{T_s} \frac{k_{\perp}^2\rho_s^2}{2} 
\approx \sum_s \frac{Z_s^2 n_s}{T_s}\frac{\vths}{\omega\dz}\left(-\frac{2 i \alpha}{\sqrt{\pi}} \sin^2\left(\frac{k_{\parl}\dz}{2}\right) + \frac{\vths}{\omega\dz}\left(2\alpha \cos(k_{\parl}\dz) \sin^2\left(\frac{k_{\parl}\dz}{2}\right) + \frac{1-\alpha}{2}\sin^2(k_{\parl}\dz)\right)\right),
\label{eqn:approx_dispersion}
\eeq
which has been truncated at second order in $1/\zeta_s$.
From~(\ref{eqn:approx_dispersion}), we see that if there is any upwinding of $\partial\phitb/\partial z$ (i.e., $\alpha > 0$), the term proportional to $1/\omega^2$ can be neglected.  The result is a spurious damped mode with maximum damping rate $\gamma$ at the grid scale in $z$ given by
\beq
\gamma \approx -\frac{4\alpha }{\sqrt{\pi}}\frac{\vthe}{\dz} \left(\sum_s Z_s^2\frac{n_s}{n_e}\frac{T_e}{T_s} k_{\perp}^2\rho_s^2\right)^{-1}.
\eeq
If $\partial\phitb/\partial z$ is discretized with centered differences (i.e., $\alpha=0$), one must retain the term proportional to $1/\omega^2$.  In this case, one recovers a discretized form of the electrostatic Alfv\'en wave:
\beq
\omega^2 \approx \sin^2(k_{\parl}\dz) \frac{\vthe^2}{\dz^2} \left(\sum_s Z_s^2\frac{n_s}{n_e}\frac{T_e}{T_s} k_{\perp}^2\rho_s^2\right)^{-1}.
\eeq
The requirement that these frequencies be resolved by an explicit time advance method leads to a severe constraint on the allowed time step size.

\bibliographystyle{elsarticle-num} 
\bibliography{general}





\end{document}